# Magnetic, electronic and Shubnikov-de Haas investigation of the dense Kondo system $CeAgSb_2$

E. Jobiliong, J. S. Brooks, E. S. Choi

*Physics/NHMFL, Florida State University, Tallahassee FL 32310*

Han-Oh Lee, Z. Fisk

*Physics, University of California at Davis, CA 95616*

Of the dense Kondo materials in the class $CeTSb_2$ (where T = Au, Ag, Ni, Cu, or Pd), $CeAgSb_2$ is special due to its complex magnetic ground state, which exhibits both ferro- and anti-ferromagnetic character below an ordering temperature $T_O \sim 9.8$ K. To further elucidate a description this magnetic ground state, we have carried out a systematic study of single crystalline $CeAgSb_2$ by magnetic, electrical magneto-transport, and Shubnikov-de Haas (SdH) studies over a broad range of temperature and magnetic field. We have constructed the magnetic phase diagram based solely on magnetoresistance data. Here, depending on the orientation of the magnetic field H, either ferromagnetic or antiferromagnetic ordering occurs below $T_O$. The resistivity of this compound below $T_O$ does not follow a simple Fermi liquid behavior, but requires an additional contribution from conduction electron scattering from boson excitations with an energy gap, $\Delta$. At zero field the temperature dependent resistivity below $T_O$ is most consistent with antiferromagnetic order, based on the transport theory which includes magnon scattering. With increasing magnetic field, the antiferromagnetic gap is observed to decrease, a result that is consistent with both field dependent magnetization and magnetoresistance data. Crystal field effect theory applied to the susceptibility data yields splitting energies from the ground state to the first and second excited states of 53 K and 137 K, respectively. Although there is some uncertainty in the Kondo temperature determination, we estimate $T_K \sim 23$ K from our analysis. In the Fermi surface studies, the measurements show very small Fermi surface sections, not predicted by band structure calculations, and the SdH amplitudes are very sensitive to field direction. Only by considering lens orbits between the main Fermi surface cylinders can the SdH results be reconciled with the Fermi surface topology predicted from band structure.

Introduction

Cerium intermetallic compounds exhibit a variety of phenomena such as heavy fermion, Kondo insulating, anisotropic transport and magnetic ordering behavior.[1,2,3] Of current interest in this class of materials are Ce compounds in the tetragonal $ZrCuSi_2$ structure (P4/nmm) including $CeNiSb_2$, $CeCuSb_2$, $CePdSb_2$, $CeAuSb_2$ and $CeAgSb_2$. These compounds exhibit competition between the Ruderman-Kittel-Kasuya-Yosida (RKKY) interaction and the Kondo effect, which leads to either magnetic or non-magnetic ground states depending on the strength of the magnetic exchange interaction $J_{cf}$ between the conduction electrons and localized 4*f*-spins.[4] The crystalline electric field (CEF) also plays a significant role in determining their magnetic properties. The CEF analysis provides important information about the hybridization effect. In addition, as proposed by Levy and Zhang, the CEF potential depends on the hybridization between the conduction band states and the localized *f*-electron states, which are responsible for the heavy fermion behavior.[5]

Of particular interest is $CeAgSb_2$, following the report of weak ferromagnetic order in polycrystalline samples by Sologub et al. [6] Here magnetization measurements on $CeAgSb_2$ showed a transition below 12 K with a net ferromagnetic moment of 0.15 $\mu_B$/Ce at 5 K. $CeAgSb_2$ crystallizes in the primitive tetragonal $ZrCuSi_2$ structure (*P*4/*nmm*), which consists of Sb-CeSb-Ag-CeSb-Sb layers along [001] with lattice constants $a$ = 4.363 A and $c$ = 10.699 A, as shown in FIG. 1.[1,7] Several different groups have investigated the magnetic properties of $CeAgSb_2$ with conflicting results for the interpretation of the magnetic ground state.[3,4,8,9,10,11] In particular, for single crystal samples, the magnetization of $CeAgSb_2$ is anisotropic[1]. In this case the magnetization for the magnetic field applied parallel to the *c*-axis shows a typical ferromagnetic signature with a saturation moment of 0.37$\mu_B$/Ce above 0.025T at 2 K. However, at the same temperature, the magnetization for the in-plane direction increases up to 3 T and then saturates with a moment of 1.2$\mu_B$/Ce. The observation of this relatively large saturated moment implies that the 4$f$ electrons are almost localized at low temperature. The structure in the magnetization at 3 T for in-plane field is indicative of a change in the antiferromagnetic spin alignment. Muro et al.[3] suggested a ferrimagnetic ground state in polycrystalline samples with a spin-flip field of about 1.3 T. However, this result disagrees with muon spin rotation $\mu$SR measurements where spectra in both the ordered state and paramagnetic state indicate a single crystallographic and magnetic muon site.[11] Neutron powder diffraction measurements indicate that the magnetic moment is oriented along the *c*-axis with a Curie temperature about 9.5 K, but with a smaller saturation moment of $0.33\mu_B$/Ce.[12] Nevertheless in accordance with $\mu$SR measurements[11], it is very difficult in polycrystalline samples to differentiate between a simple ferromagnetic structure and a complex antiferromagnetic structure with a resultant ferromagnetic component. Inelastic neutron scattering[8] and $\mu$SR experiments[11] indicates a relatively high Kondo temperature of 60 to 80 K and 60 K, respectively. However, the magnetic entropy reaches almost $R$ln2/(mol Ce) at the ordering temperature $T_O$,[1] suggesting the Kondo temperature may be closer to $T_O$. Complex features indicating anisotropy are also evident in electronic transport measurements. The transverse magnetoresistance (MR) of $CeAgSb_2$ is found to be strongly anisotropic. The MR for the field parallel to the *c*-axis is monotonic, but for field in the basal plane, a kink appears near 3 T corresponding to the saturation in the magnetization for the same field direction.[1]

Recent measurements of the de Haas-van Alphen (dHvA) effect[2] have been used to study the Fermi surface of this compound. The electronic specific heat coefficient $\gamma$ of single crystal[13] $CeAgSb_2$ is 65 mJ $K^{-2}mol^{-1}$ (75 mJ $K^{-2}mol^{-1}$ for polycrystalline samples[3]) indicative of heavy mass carriers. Effective cyclotron masses from dHvA measurements for fields along the *c*-axis are between $0.85m_e$ ($m_e$ is free electron mass) and $32m_e$, for dHvA frequencies between 41 T and 11.2 kT.[2] The band-structure calculations predict that the Fermi surface of $CeAgSb_2$ has a large dHvA frequency of 10.7 kT and several branches with the dHvA frequencies between 4 kT and 9 kT.[2] Only one previous Shubnikov-de Haas (SdH) measurement has been reported for magnetic field parallel to *c*-axis. Here, a single orbit of ~ 25 T has been observed at 1.2 kbar in the range 18 T at 2.1 K.[14]

The purpose of the present work has been to investigate in more detail some aspects of single crystal $CeAgSb_2$ samples to better determine the assignment of a magnetic ordering in the ground state, and to further explore the behavior of the Fermi surface through the quantum oscillations. In particular, detailed studies of the temperature dependent magnetoresitance can be described by terms involving scattering due to magnons in addition to normal Fermi liquid behavior. This leads to a description of an anisotropic magnetic ground state, where ferro- and anti-ferromagnetic order depends on magnetic field direction. We find that the magnetization and susceptibility are well described by CEF theory where the energy level parameters are consistent with inelastic neutron scattering experiments. In addition, we find, as in a previous report, that one quantum oscillation dominates the SdH spectrum, which is smaller in extremal area than any oscillation seen in the previous dHvA study. Angular dependent SdH studies further reveal unusual aspects of the Fermi surface of this compound.

**Experimental**

Single crystals of $CeAgSb_2$ were made with excess Sb as a flux. The starting materials were placed in an alumina crucible and sealed under vacuum in a quartz ampule, heated to 1150 C, and then cooled slowly to 670 C and centrifuged to remove the flux. The DC resistivity data were measured using a conventional four-probe method with current applied in *ab* –plane. The typical size of a single crystal is 2.5 mm × 1 mm × 0.3 mm.

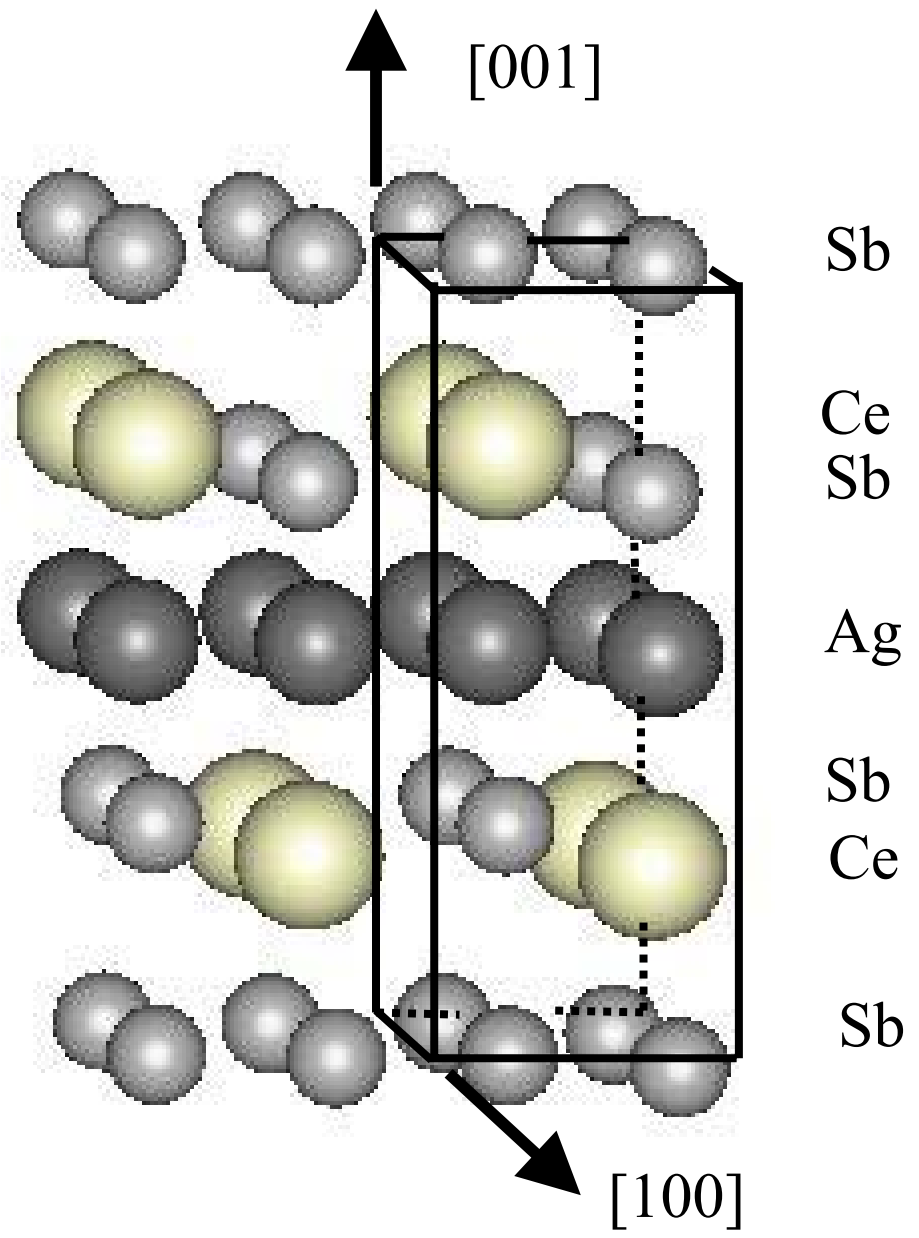

FIG. 1 Crystal structure of $CeAgSb_2$. The tetragonal volume defines the unit cell.

The magnetization studies were carried out in a Superconducting Quantum Interference Device (SQUID) magnetometer over the temperature range 1.8 – 300 K in the field range 0 to 5.5 T. Shubnikov-de Haas measurements were performed in both a 33 T resistive magnet in a helium cryostat, and separately in a 18 T superconducting magnet with a dilution refrigerator.

**Results and Analysis**

**1. Electrical transport properties**

The DC resistivity of a single crystal of $CeAgSb_2$ vs. temperature in the range room temperature to ~300 mK is shown in FIG. 2. At high temperature the scattering is phonon-dominated and the resistivity decreases with decreasing temperature. However, below ~150 K, the resistivity increases logarithmically as the temperature decreases, characteristic of a Kondo lattice system. Below ~15 K the resistivity exhibits behavior marked by drop in resistivity where there is an onset in coherent scattering in the Kondo lattice, as denoted by $T_{coh}$ (see the inset of FIG. 2). Such a low $T_{coh}$ is consistent with most of the Ce compounds and suggests the system is weakly hybridized.[15] We are able to fit the temperature dependence of the resisitivity in the range ~300 K to ~30 K, as shown in the inset of FIG. 2, by using[16]

$$\rho(T) = \rho_0^\infty - C_1 \ln T + C_2 T^5 \int_0^{\Theta_D/T} \frac{x^5}{\left(e^x - 1\right)\left(1 - e^{-x}\right)} dx, \quad (1)$$

where $\rho_0^\infty$ is the resistivity due to spin-disorder, $C_1$ is the Kondo coefficient, $C_2$ is a temperature-independent constant related to electron-phonon strength and $\Theta_D$ is the Debye temperature. The second and third term of Eq.1 describe the characteristic Kondo lattice and the electron-phonon scattering, also known as the Bloch-Grüneisen relation, respectively. We find the parameters corresponding to this fit are $\rho_0^\infty$ = 251 μΩ-cm, $C_1$ = 33.8 μΩ-cm, $C_2 = 3.08\times10^{-10}$ μΩ-cm $K^5$, $\Theta_D$ = 216 K. For comparison, $\Theta_D$ is ~ 200 K for non-magnetic $LaAgSb_2$, estimated from specific heat measurements.[3,4] Below about 10 K, (see the inset of FIG. 2), the resistivity decreases significantly, corresponding to a magnetic transition at $T_O$ ~9.8 K. The residual resistivity $\rho_0$ (T ~ 300 mK) and the residual resistivity ratio (RRR) ($=\rho_{room\ temp}/\rho_0$) are 0.126 μΩ-cm and 853, respectively, reflecting the high quality of the sample.

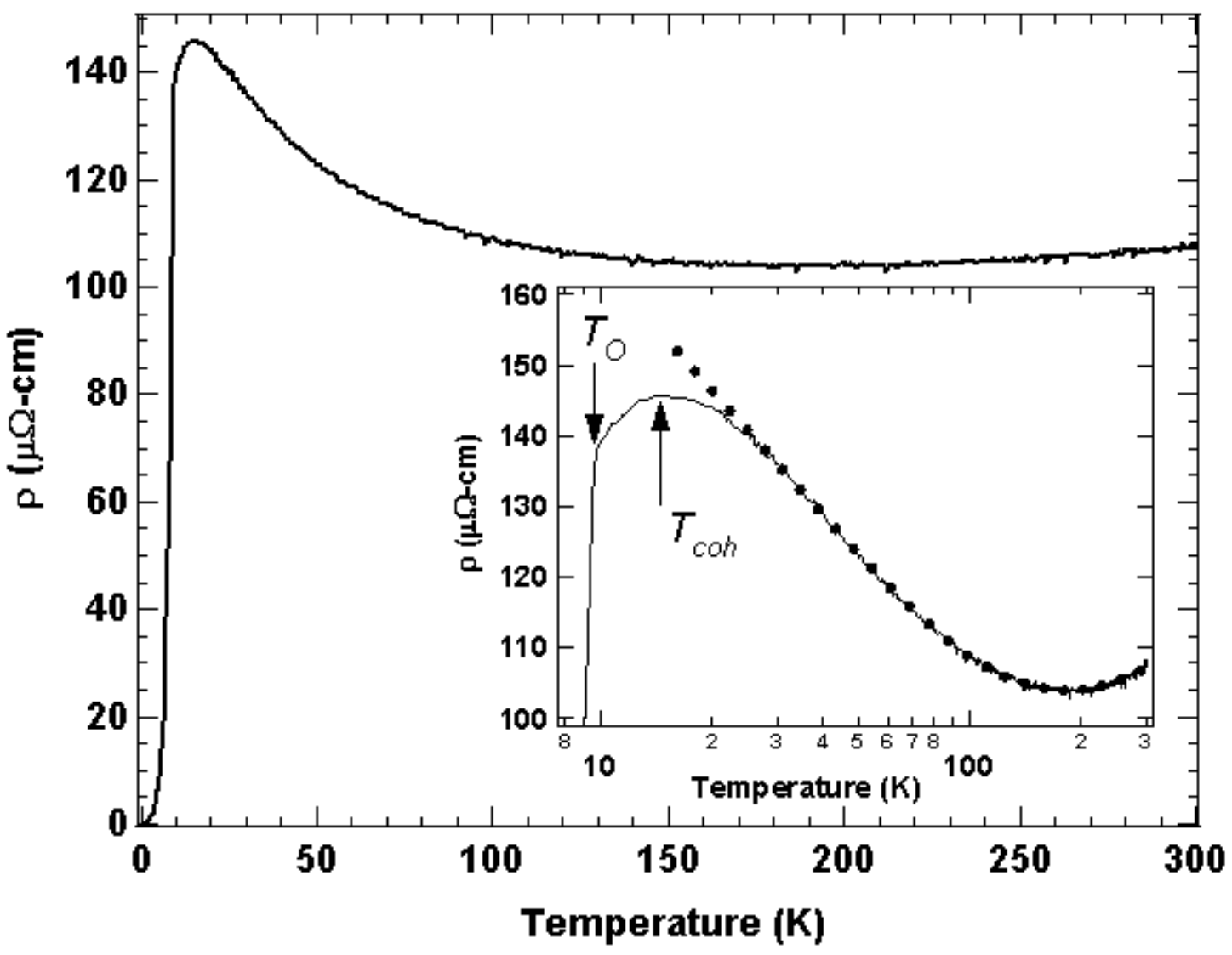

FIG. 2 Temperature dependence of the in-plane DC resistivity of a single crystal of $CeAgSb_2$ between room temperature and ~ 300 mK. Inset: High temperature fit (dotted line) to Eq.1.

The temperature dependence of the in-plane DC resistivity below $T_O$ does not follow a simple Fermi-liquid behavior ($\rho \sim \rho_0 + AT^2$), as shown in FIG. 3 by the solid line, but has an additional temperature dependence. The additional term takes into account the resistivity due to electron-magnon scattering. In general, the resistivity due to electrons scattering from an arbitrary type of boson excitation (magnon or phonon) can be written as

$$\rho_B = \left(\frac{m}{ne^2}\right)\frac{1}{\tau}, \tag{2}$$

where $n = k_F^3 / 3\pi^2$ is the number density and the scattering time $\tau$ is given by[17]

$$\frac{1}{\tau} = \pi N(0)\int_0^{2k_F} \frac{1}{k_F^2} k^3 dk \int \frac{d\Omega_{\vec{k}}}{4\pi} \times \left|g_{\vec{k}}\right|^2 \times \frac{\hbar\omega_{\vec{k}} / k_B T}{4Sinh^2\left(\hbar\omega_{\vec{k}} / 2k_B T\right)}. \tag{3}$$

Here, $N(0) = mk_F / 2\pi^2\hbar^2$ is the density of states per spin at the Fermi level, $2k_F$ represents the maximum wave-vector transfer, $g_{\vec{k}}$ is the electron-boson coupling, $k_B$ is the Boltzmann constant and $\hbar\omega_{\vec{k}}$ is the boson energy for a given wave vector $\vec{k}$ . We may apply Eq.3 to either the ferromagnetic or to the antiferromagnetic case, as discussed in the following sections.

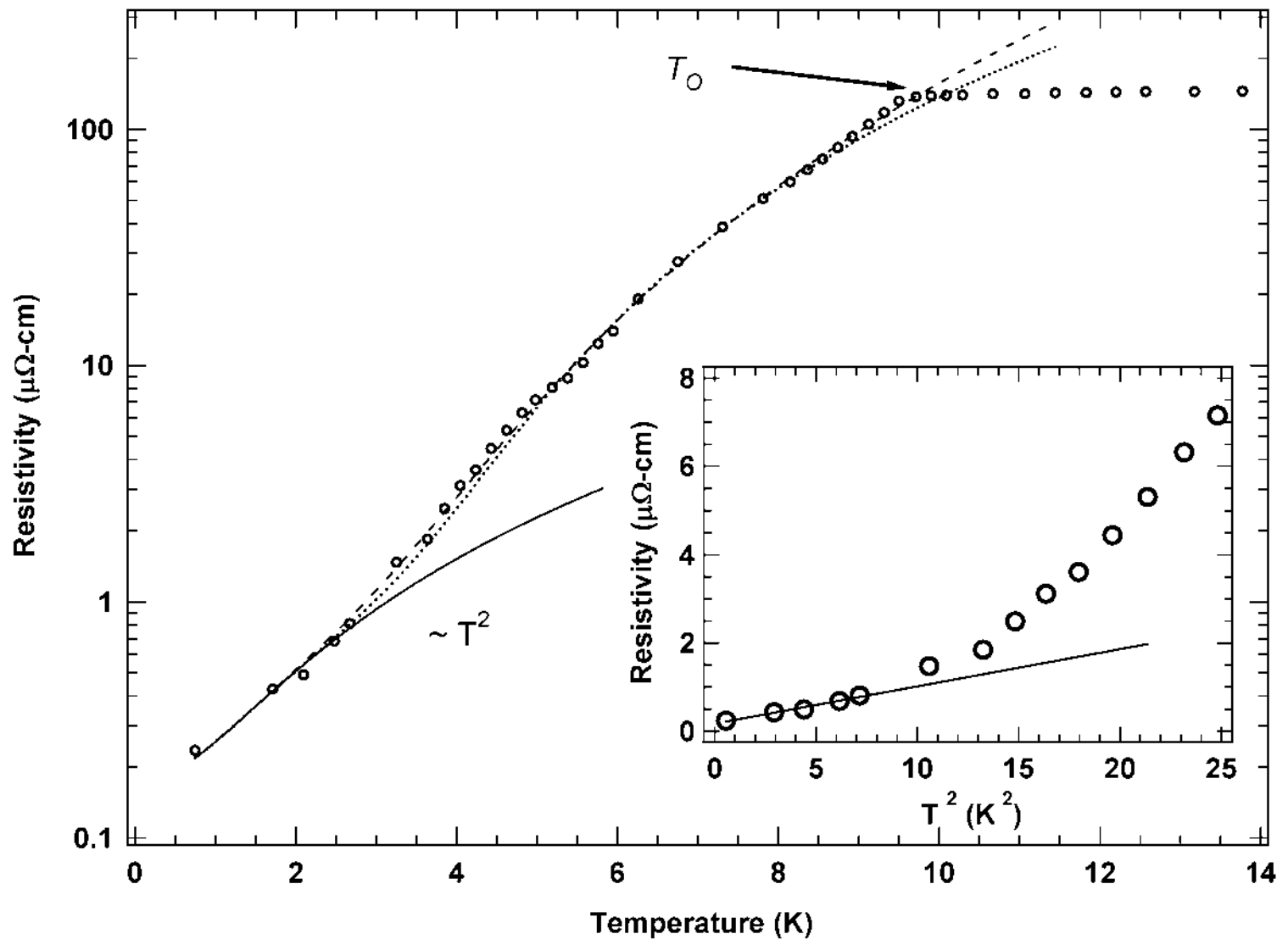

FIG. 3 Temperature dependence of the in-plane DC resistivity of a single crystal of $CeAgSb_2$ at low temperature. The solid line indicates $T^2$ dependence and the dashed and dotted line indicate the full fit to Eq.11 for AFM and FM, respectively. Inset: Low-temperature fit of $\rho(T) = \rho_0 + AT^2$.

## 1.1 Ferromagnetic case

In the case of an anisotropic ferromagnetic (FM) material, there is a gap Δ in the magnon spectrum, and the energy dispersion relation of the magnon[17] can be expressed by $\hbar\omega_{\vec{k}} = \Delta + C_0 k^2$, where $C_0$ is the spin wave stiffness. Since the electron-magnon coupling $\left|g_{\vec{k}}\right|^2$ for ferromagnetic system is independent[18] of $\vec{k}$, one may use the approximation

$$\frac{1}{4Sinh^2\left(\hbar\omega_{\vec{k}} / 2k_B T\right)} \approx e^{-\hbar\omega_{\vec{k}}/k_B T}. \tag{4}$$

Thus, the electron-magnon resistivity in the anisotropic ferromagnetic material $\rho_{FM}$ can be obtained by using the energy dispersion relation, constant electron-magnon coupling, and Eq.3. The leading term of the resistivity in the ferromagnetic case is

$$\rho_{FM}\left(T\right) = BT\Delta\left(1 + \frac{2T}{\Delta}\right)e^{-\Delta/T}, \tag{5}$$

where $B$ is a constant related to the spin disorder scattering.[17]

### 1.2 Antiferromagnetic case

In the case of antiferromagnetic (AFM) system, the electron-magnon coupling[17] $\left|g_{\vec{k}}\right|^2 \propto k$ and the energy dispersion relation[19] is $\hbar\omega_{\vec{k}} = \sqrt{\Delta^2 + Dk^2}$ . By using Eqs. 1 and 2, along with the information given above, we obtain

$$\rho_{AFM}\left(T\right) = \frac{C}{T}\int_0^{\infty} \frac{k^4\sqrt{\Delta^2 + Dk^2}}{Sinh^2\left(\sqrt{\Delta^2 + Dk^2}\,/\,2T\right)}dk, \tag{6}$$

where $C$ is a constant. Using the change of variable, $y = \frac{\sqrt{\Delta^2 + Dk^2}}{2T}$, one can approximate $Sinh^2 y \approx e^{-2y}$ in the small temperature limit $T \ll \Delta$, we find

$$\rho_{AFM}\left(T\right) \approx CT^5\int_{y_0}^{\infty} y^2\left(y^2 - y_0^2\right)^{3/2} e^{-2y}dy, \tag{7}$$

where $y_0 = \frac{\Delta}{2T}$. This integral can be easily calculated if we use another change of variable $y = y_0 Cosh(x)$, so that

$$\rho_{AFM}\left(T\right) = CT^5 y_0^5 e^{-2y_0}\int_0^{\infty} Cosh^2 x\ Sinh^3 x e^{-2y_0 Cosh\,x + 2y_0}dx. \tag{8}$$

In the limit of $y_0 = \frac{\Delta}{2T} \ll 1$, the integrand of Eq.8 is asymptotic to zero above a certain $x$, so that it can be simplified to $Cosh^2 x\ Sinh^3 x$, in which the function $e^{-2y_0 Cosh\,x + 2y_0}$ acts as a cutoff in the integral. The limit of the integral then is from 0 to $x_C$, where $x_C$ is the solution of $-2y_0 Cosh\ x_C + 2y_0 = -1$, and Eq.8 becomes

$$\rho_{AFM}\left(T\right) = CT^5 y_0^5 e^{-2y_0}\int_0^{x_C} Cosh^2 x\ Sinh^3 x dx. \tag{9}$$

Thus, the leading term of the resistivity in the antiferromagnetic case is given by

$$\rho_{AFM}\left(T\right) \approx C\Delta^5 e^{-\Delta/T}\left\{\frac{1}{5}\left(\frac{T}{\Delta}\right)^5 + \left(\frac{T}{\Delta}\right)^4 + \frac{5}{3}\left(\frac{T}{\Delta}\right)^3\right\}. \tag{10}$$

### 1.3 Magnetotransport analysis

The total resistivity of this compound below the transition temperature can be written as

$$\text{FM-case:} \quad \rho_{FM}\left(T\right) = \rho_0 + AT^2 + BT\Delta\left(1+\frac{2T}{\Delta}\right)e^{-\Delta/T} \text{ , or}$$

$$\text{AFM-case:} \quad \rho_{AFM}\left(T\right) = \rho_0 + AT^2 + C\Delta^5 e^{-\Delta/T}\left\{\frac{1}{5}\left(\frac{T}{\Delta}\right)^5 + \left(\frac{T}{\Delta}\right)^4 + \frac{5}{3}\left(\frac{T}{\Delta}\right)^3\right\}. \tag{11}$$

In the gapless limt $\Delta \rightarrow 0$, we obtain $\rho \sim T^2$ and $\rho \sim T^5$ from Eq.11 for ferromagnetic and antiferromagnetic case, respectively, as discussed in Ref.17.

In practice, to apply Eq.11 to our data, we fit the first two terms of Eq. 11 at very low temperature (0.7 K to ~3 K) to obtain $\rho_0$ and $A$, as shown in the inset of FIG. 3. These parameters are then fixed and the parameters in Eq.11 are obtained by a higher temperature fit up to ~8K for both FM and AFM cases. We find the parameters corresponding to this fit are $\rho_0$ = 0.171 μΩ-cm, $A$ = 0.0845 μΩ-cm.K$^{-2}$; $B$ = 3.3K$^{-2}$, $\Delta$ = 24.3 K (FM) and $C$ = 1.3×10$^{-3}$ μΩ-cm.K$^{-5}$, $\Delta$ = 10.5 K (AFM). Although small, there is a significant difference between these two fits at lower temperatures and fields. By applying this fitting procedure to the field dependent resistivity for H ⊥ $c$-axis, we are able to see more clearly these differences, as shown in FIG. 4a, where in the low field limit, the AFM description gives the best fit. This suggests antiferromagnetic ordering in the basal plane below $T_O$. At higher fields, the spins will align parallel to the external magnetic field, favoring ferromagnetic order. We have determined the effect of the magnetic field on the AFM energy gap $\Delta_{AFM}$, as shown in FIG. 4b, and find the magnetic field reduces the gap energy. This is not surprising because in the antiferromagnetic case, the gap will be modified[20] by $\Delta_H = \Delta - \mu_{eff} H_{eff}$, where $\mu_{eff}$ is the effective magnetic moment and $H_{eff} = H - H_M$ which is the sum of the applied magnetic field and molecular field $H_M$ produced by the other moments. In contrast, for H // $c$-axis both FM and AFM fits work well at low fields, but only the FM fit is adequate at higher fields. Thus, this suggests that the ferromagnetic ordering is along the $c$-axis.

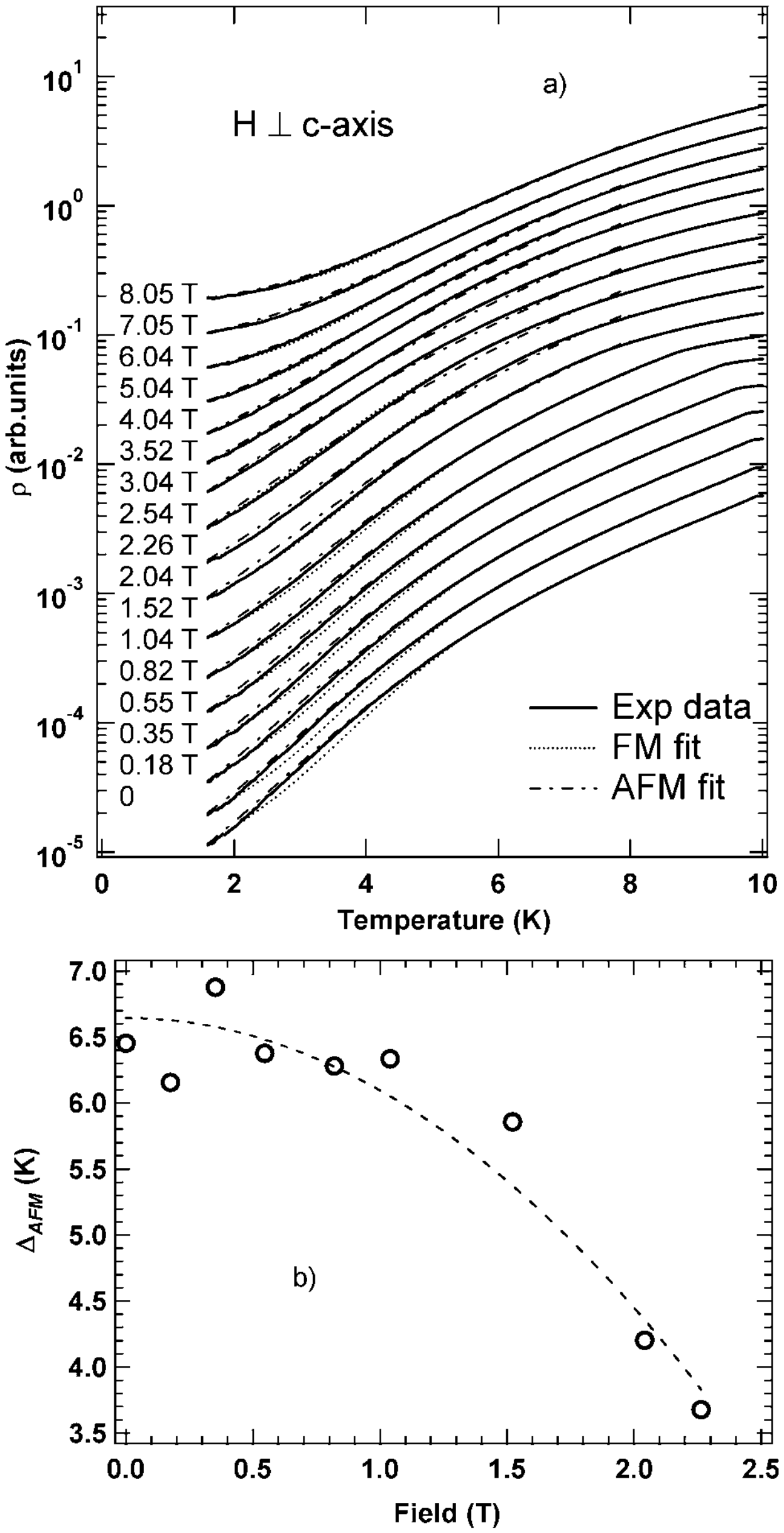

FIG. 4 a) The in-plane DC resistivity (curves offset for clarity) vs. temperature for different fields (H ⊥ *c*-axis). The dotted and dashed-dotted lines indicate fits to FM and AFM, respectively. b) The field dependence of AFM gap energy, $\Delta_{AFM}$. The dashed line is a guide to the eye.

The effect of magnetic field on the ordering temperature $T_O$ is shown, for both magnetic field applied parallel and perpendicular to the *c*-axis in FIG. 5. We use the minimum of the second derivative of the resistivity with respect to the temperature ($\frac{d}{dT}\left(\frac{d^2\rho}{dT^2}\right)=0$) as the criterion for the ordering temperature, which is indicated by an arrow in FIG. 5. We observed that the ordering temperatures have different behavior depending on the direction of the applied magnetic field. For H ⊥ *c*-axis, $T_O$ decreases as the external field increases, following antiferromagnetic behavior. In contrast, for H //*c*-axis, $T_O$ increases as the external field increases, which is consistent with ferromagnetic order. From this analysis, we are able to construct the magnetic phase diagram of this compound, as shown in FIG. 6.

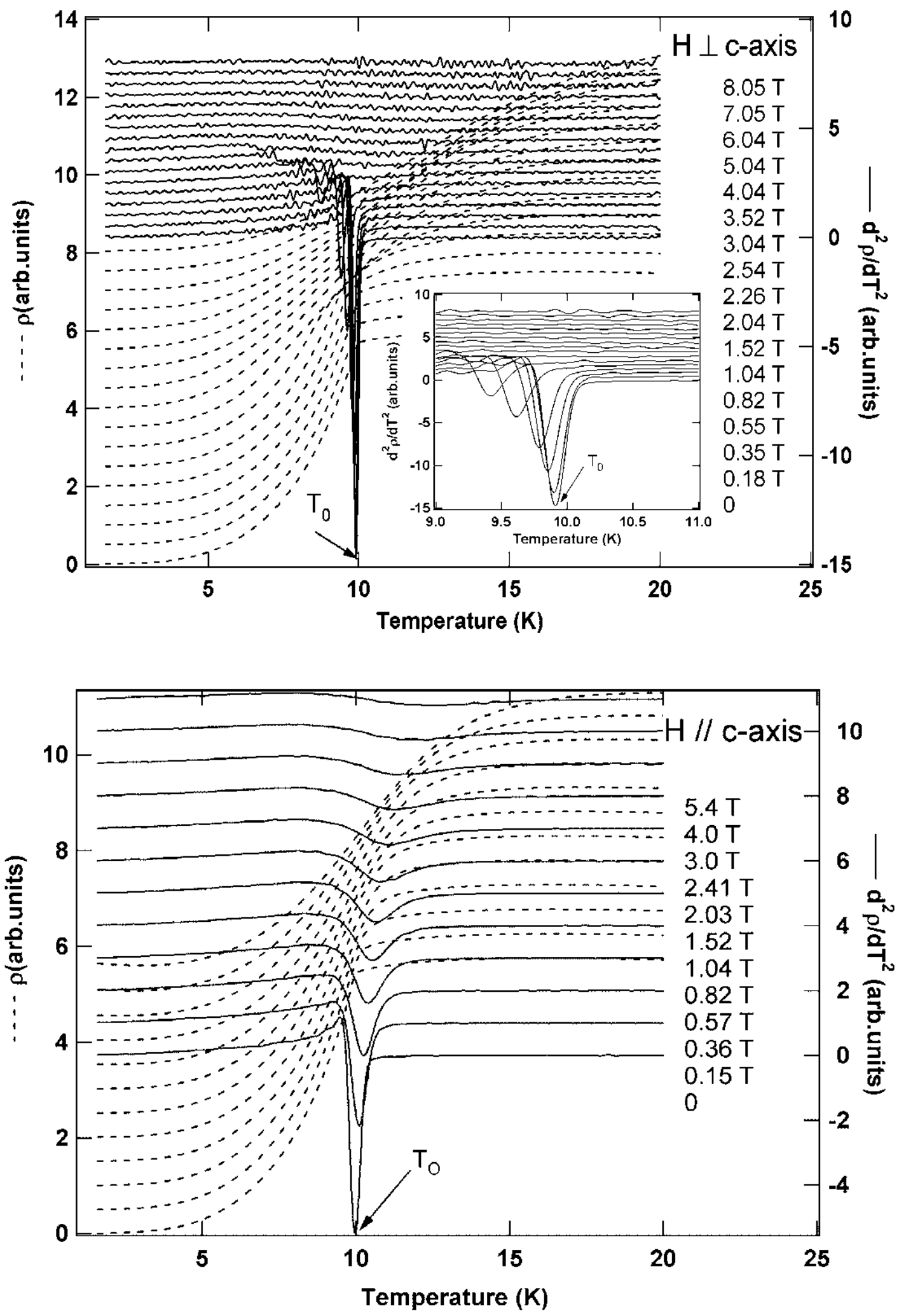

FIG. 5 The in-plane DC resistivity (curves offset for clarity) vs. temperature for different fields (top: H ⊥ *c*-axis ; bottom: H // *c*-axis). The dashed and solid lines indicate the experimental data and $d^2\rho/dT^2$, respectively. The arrow indicates the transition temperature $T_O$. Inset: expanded view of $d^2\rho/dT^2$ around $T_O$ for H ⊥ *c*

This result is similar to that obtained in Ref. 4, estimated from the magnetization and the thermal expansion measurements. We will see below that this magnetic phase diagram is consistent with the crystalline electric field (CEF) analysis.

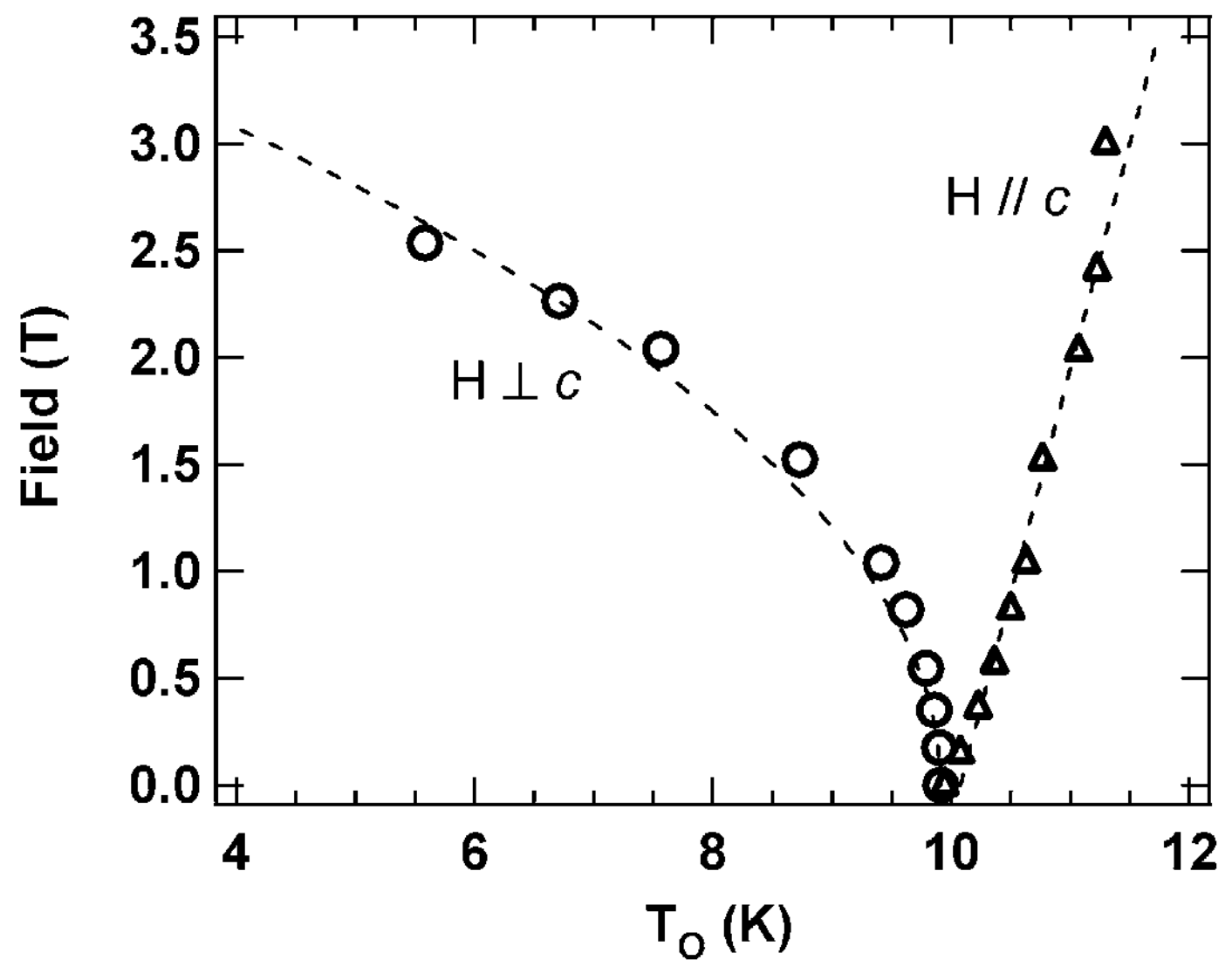

FIG. 6 Magnetic phase diagram of $CeAgSb_2$ based on the MR data. The open circles and triangles are the transition field for H ⊥ *c*-axis and H // *c*-axis, respectively, obtained from FIG. 5. The dashed lines are guides to the eye.

The magnetoresistance (MR) of this compound, defined as $\{\rho(H,T)-\rho(0,T)\}/\rho(0,T)$, for the two different orientations is shown in FIG. 7. The MR changes sign below a characteristic temperature $T_m$ from negative to positive, where $T_m$ also decreases or increases depending on the direction of the applied field (see the insets in FIG. 7). The anisotropy of $T_m$ also suggests that the system has different magnetic ordering for the different field directions. In both cases, $T_m$ saturates at a certain field. The behavior of the MR above and below $T_m$ can be explained as follows. Above $T_m$, the negative character of the magnetoresistance is due to the reduction in electron-spin scattering, where as the magnetic field increases, the effective field suppresses the fluctuations of the localized spins, leading to an increase in the conductivity. Below $T_m$, when the magnetic field increases, the gap energy Δ decreases and more magnons will be in the excited state, which causes more electron-magnon scattering, increasing the resistivity.

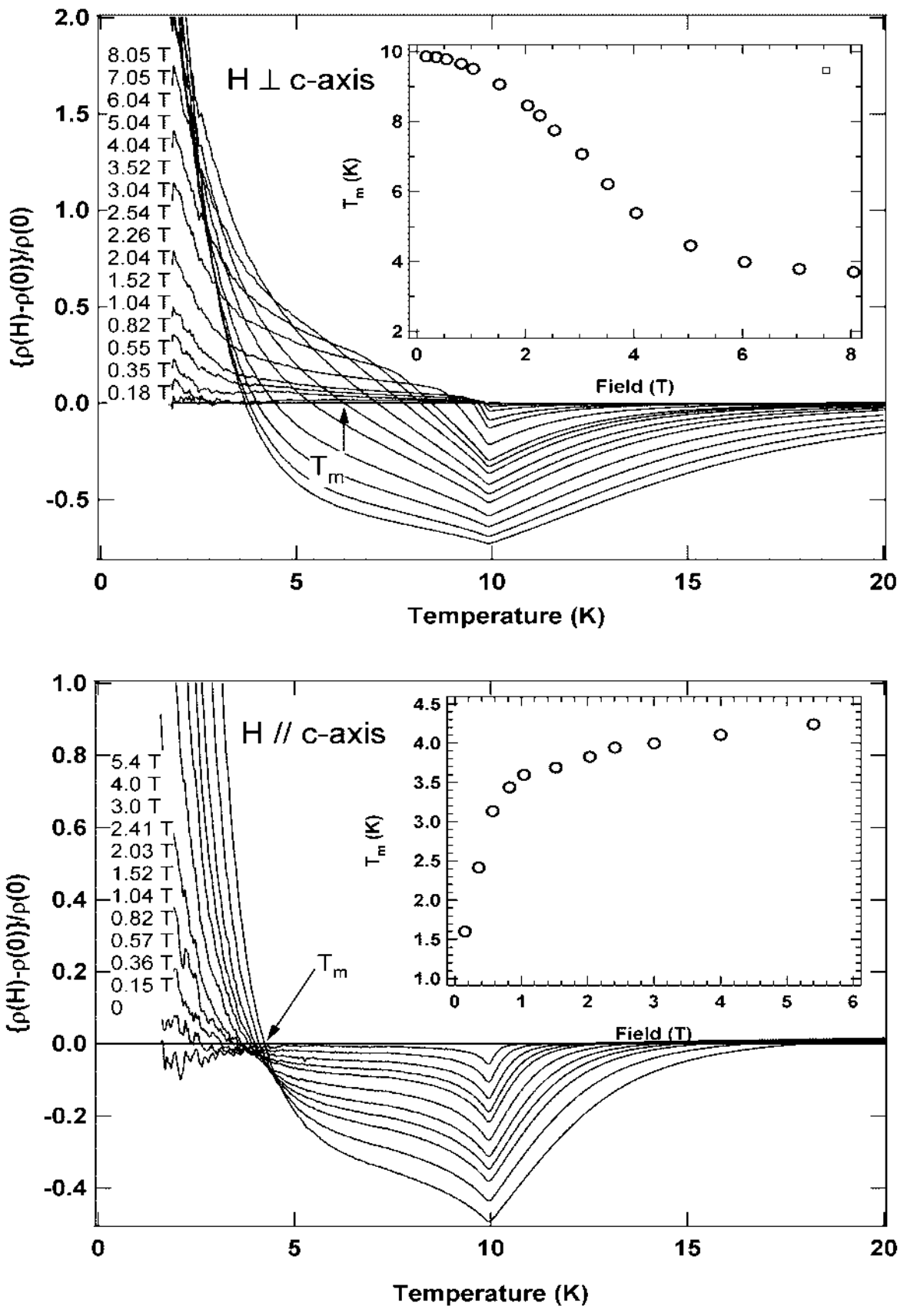

FIG. 7 The magnetoresistance of $CeAgSb_2$ as a function of temperature for top: H $\perp$ *c*-axis and bottom: H // *c*-axis. Inset: The field dependence of the characteristic $T_m$ at which $[\rho(H)-\rho(0)] = 0$.

## 2. Magnetic properties

The magnetization measurement of $CeAgSb_2$ exhibits magnetic ordering below $T_O$ ~9.8 K as shown in FIG. 8a, where we note that below $T_O$ the magnetization is anisotropic with respect to field direction. The temperature dependence of the magnetization under 0.1 T for field perpendicular to the *c*-axis shows a cusp around $T_O$, which is usually found in an antiferromagnetic transition. We have used the Curie-Weiss law

$$\chi = \frac{C}{T-\Theta_C} + \chi_0, \tag{12}$$

to fit the susceptibility data. Here $C = \frac{N\mu_{eff}^2}{k_b}$ and $\chi_0$ are the Curie constant and temperature-independent susceptibility, respectively. For H//*c* the effective magnetic moment and Curie temperature are $\mu_{eff}$ = 2.51$\mu_B$/Ce and $\Theta_C$ = -63.9 K. For H⊥*c*, $\mu_{eff}$ = 2.48$\mu_B$/Ce and $\Theta_C$ = 5.05 K. Both effective magnetic moments are close to the theoretical value of 2.54$\mu_B$/Ce for $Ce^{3+}$ (S= ½ ; L=3 ; J=5/2).

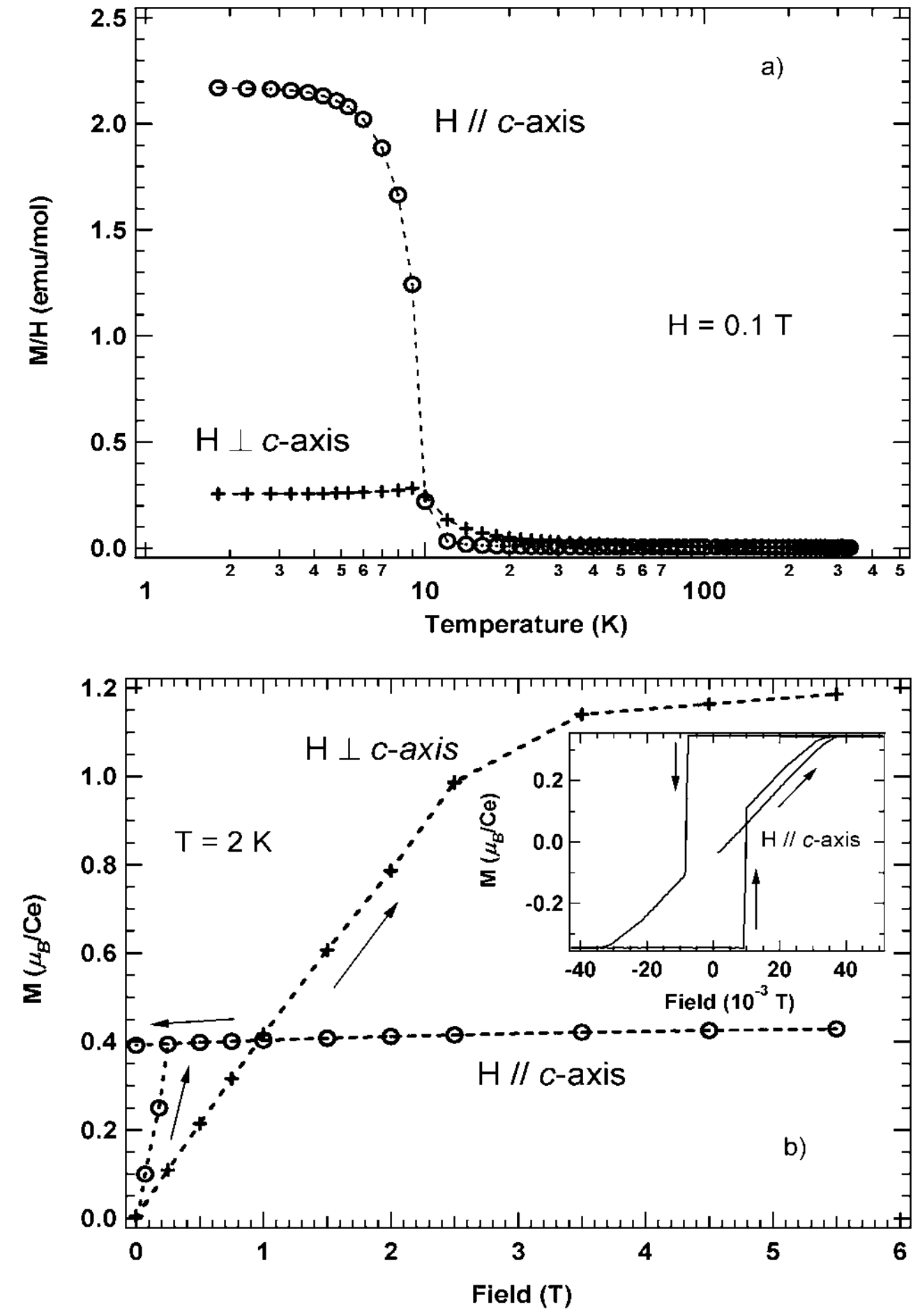

FIG. 8 a) Temperature dependence of the susceptibility of $CeAgSb_2$ both for field applied parallel and perpendicular to the *c*-axis, at H=0.1T. b) Field dependence of magnetization for two different field directions at T=2 K. The dashed lines are guides to the eye. Inset: The magnetization at low field for H//*c*-axis.

The magnetization isotherm for H⊥*c* data at T = 2 K, as shown in FIG. 8b, increases almost linearly below 3.5T and then remains nearly constant at high field with a saturated moment ~1.2$\mu_B$/Ce. No hysteresis is found, suggesting this compound is antiferromagnetically ordered for H⊥*c*. In contrast, for H//*c* a saturation magnetic moment of ~0.4 $\mu_B$/Ce is found at a low field (~0.04T). Hysteresis with a remnant magnetic field ~ -0.01T is observed, as shown in the inset of FIG. 8b, indicating ferromagnetic order.

The Kondo temperature for a single-ion and non-magnetic order can be estimated by using[21,22]

$$T_K = \frac{(\nu - 1)\pi R W}{6\gamma}, \tag{13}$$

where $\nu = 2J + 1 = 2$, $W$ is the Wilson number = 1.289, $R$ is the universal gas constant = 8.314 $J.mol^{-1}.K^{-1}$ and $\gamma$ is the Sommerfeld coefficient ~ 244 $mJ.mol^{-1}.K^{-2}$. We estimate the Sommerfeld coefficient by using $\frac{C(T)}{T} = \gamma + aT^2$, where $C$ is the specific heat, in the temperature range between 20 K and 12 K, which is above the ordering temperature. The Kondo temperature obtained from Eq.13 is $T_K$ = 23 K, consistent with the transport data that shows coherent scattering around $T_{coh}$ ~ 15 K (see FIG. 2), suggesting that $T_K$ is similar to $T_{coh}$. The measured change in entropy in this compound at $T_O$ is ~1.04$R$ln2, also indicating the Kondo temperature must be comparable to $T_O$ (here, $T_K$ ~ 2$T_O$). However, such a small $T_K$ (23 K) is inconsistent with inelastic neutron scattering ($T_K$ is between 60 and 80 K)[8] and μSR ($T_K$ ~ 60K).[11] As we will discuss below, the crystal electric field strongly affects the magnetic properties and the energy splitting of the ground and first excited state. This splitting, which is 53 K (see table 1), is close to $T_K$ estimated from neutron, and $\mu$SR measurements, and may explain why the Kondo temperature is not well defined.

### 2.1 Crystalline electric field theory

We next discuss the magnetic properties based on the crystalline electric field (CEF) theory. The total Hamiltonian is given as follows,

$$\mathcal{H} = \mathcal{H}_{CEF} - g_j \mu_B J_i (H_i + \lambda_i M_i), \tag{14}$$

where $g_j$ is the Lande $g$ factor (6/7 for $Ce^{3+}$), $\mu_B$ is the Bohr magnetron, $J_i$ ($i = x, y$ and $z$) is the component of angular momentum, $M_i$ is the magnetization and $\mathcal{H}_{CEF}$ is the CEF Hamiltonian. The second and third terms of the Hamiltonian are the contributions from the Zeeman effect and the molecular field. The CEF Hamiltonian of this system, which has a tetragonal symmetry, can be written as

$$\mathcal{H}_{CEF} = B_2^0 O_2^0 + B_4^0 O_4^0 + B_4^4 O_4^4, \tag{15}$$

where $B_k^q$ and $O_k^q$ are the CEF parameters and the Stevens operators, respectively.[23,24] The $Ce^{3+}$ ($4f^1$) ion has an odd number of electrons in the $4f$ shell, qualifies as a Kramers ion with a doublet ground state. The CEF effect splits the $4f$-level into three doublets with excitation energy $\Delta_1$ and $\Delta_2$ from the ground state to the first and second excited states, respectively. The temperature dependence of the susceptibility based on CEF model can be expressed as

$$\chi_{CEF}^{i} = N(g_J \mu_B)^2 \frac{1}{Z}\left( \sum_{m\neq n} \left|\langle m|J_i|n\rangle\right|^2 \times \frac{1-e^{-\Delta_{mn}/k_BT}}{\Delta_{mn}} \times e^{-E_n/k_BT} + \frac{1}{k_BT}\sum_n \left|\langle n|J_i|n\rangle\right|^2 e^{-E_n/k_BT} \right). \quad (16)$$

Here, index $i$ indicates the axis ($x$, $y$ or $z$ axis), $N$ is the number of ions, $E_n$ is the energy at state-$n$, $Z$ is a partition function and $\Delta_{mn} = E_n - E_m$. The total magnetic susceptibility including the molecular field contribution is given by

$$\chi_i^{-1} = \left(\chi_{CEF}^{i}\right)^{-1} - \lambda_i . \quad (17)$$

FIG. 9 shows the inverse magnetic susceptibility as a function of temperature for different field directions. The calculated susceptibility for H//$c$ agrees well with the experimental results, but for H⊥$c$ there is a small deviation. We obtain the CEF parameters (see Table 1) by fitting the data using Eq.17. In this model, we find that the wave functions of the ground state are $\left|\pm\frac{1}{2}\right\rangle$ with a saturation moment of 0.4 $\mu_B$/Ce along the $c$-axis, in agreement with the predicted saturation moment of the ground state, $g_J \mu_B J_z$ = 0.43 $\mu_B$/Ce. The energy levels obtained by this fitting are consistent with previous results[4]. Also, the excitation energies of ~59K and ~144K are consistent with inelastic neutron scattering experiments[4]. Furthermore, neutron diffraction experiments suggest that the magnetic moments are oriented ferromagnetically along the $c$-axis with a value 0.33 $\mu_B$.[12] The molecular field parameter $\lambda$ is proportional to the exchange interaction between nearest neighbors and is negative and positive for antiferromagnetic and ferromagnetic case, respectively. In Table 1, we find $\lambda$ is negative for the H⊥$c$ and positive for H//$c$, which is consistent with the magnetic phase diagram obtained from resistivity measurement (see FIG. 6).

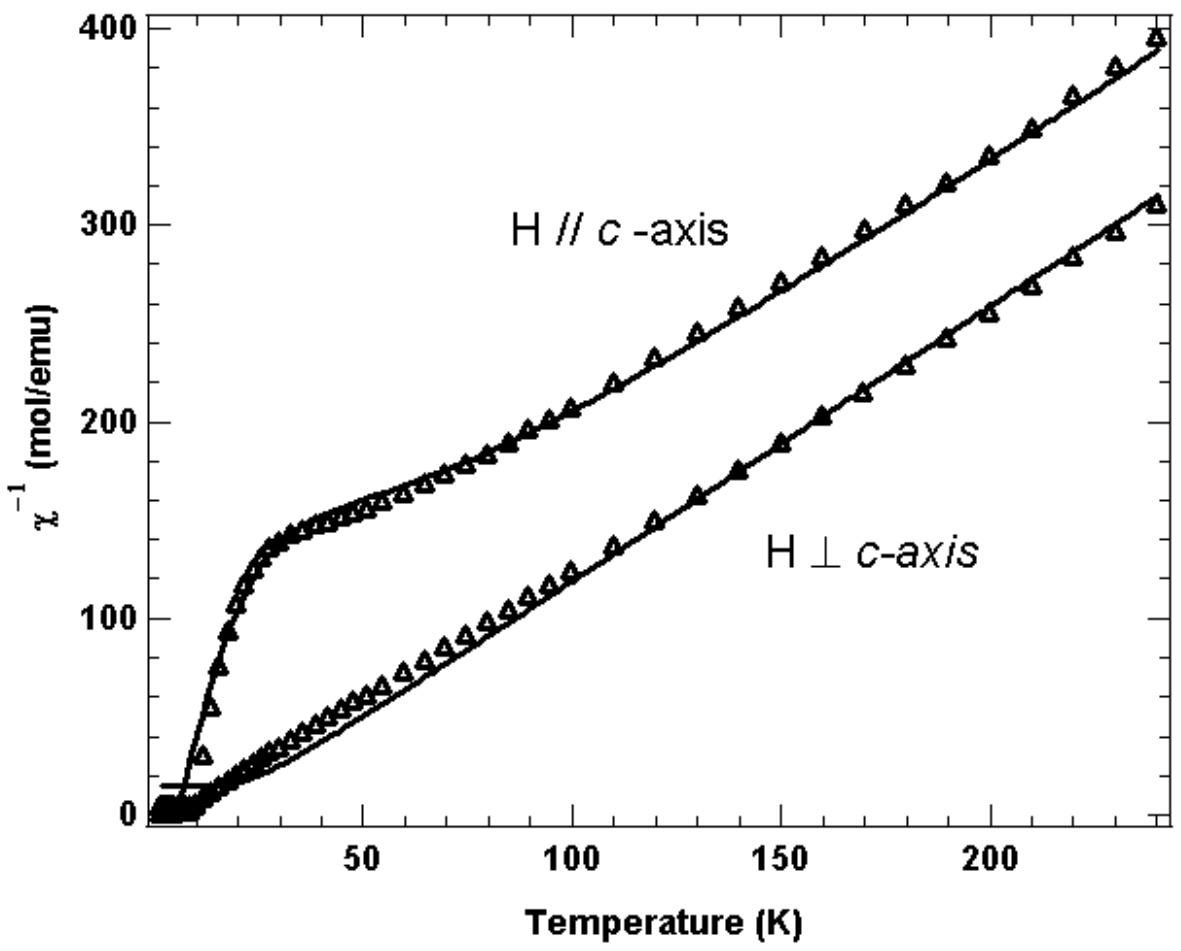

FIG. 9 The temperature dependence of the inverse magnetic susceptibility of $CeAgSb_2$ for two field directions at H = 0.1 T. Triangles indicate experimental data and the solid lines are the calculated curves based on Eq.17.

Table 1 CEF parameters, Energy level, Molecular field parameters λ and corresponding wave functions for $CeAgSb_2$

| CEF parameters | $B_2^0 = 6.60K$ | | $B_4^0 = -0.09K$ | | $B_4^4 = 1.14K$ | |
|---|---|---|---|---|---|---|
| $\lambda_{// c\text{-}axis}$ = 47.9 mol/emu | | | | $\lambda_{\perp c-axis}$ = -14.8 mol/emu | | |
| *Energy levels and wave functions* | | | | | | |
| Energy (*K*) | $\lvert +\frac{5}{2}\rangle$ | $\lvert +\frac{3}{2}\rangle$ | $\lvert +\frac{1}{2}\rangle$ | $\lvert -\frac{1}{2}\rangle$ | $\lvert -\frac{3}{2}\rangle$ | $\lvert -\frac{5}{2}\rangle$ |
| 137 | 0.919 | 0 | 0 | 0 | -0.394 | 0 |
| 137 | 0 | 0.394 | 0 | 0 | 0 | -0.919 |
| 53 | 0 | -0.919 | 0 | 0 | 0 | -0.394 |
| 53 | 0.394 | 0 | 0 | 0 | 0.919 | 0 |
| 0 | 0 | 0 | 1 | 0 | 0 | 0 |
| 0 | 0 | 0 | 0 | 1 | 0 | 0 |

**3. Shubnikov-de Haas oscillations**

The Fermi surface of $CeAgSb_2$ has been systematically studied by Inada et al[2] by angular dependent de Haas –van Alphen (dHvA) measurements, which involve the determination of the oscillatory magnetization vs. inverse field. The large cylindrical corrugated Fermi surfaces with large cyclotron masses (20-30 $m_e$) were obtained in Ref.2. Theoretically, the modified 4*f*-localized electron band calculation was proposed to explain some of the dHvA frequencies observed in the experiment. We have studied the temperature dependence of the Shubnikov-de Haas oscillations (SdH) in $CeAgSb_2$ in high fields (to 32 T) to helium temperatures, and in lower fields (18 T) to 80 mK vs. magnetic field direction. In the SdH measurement the oscillatory magnetoresistance, upon which may be imposed a background magnetoresistance, is measured. In a previous study[14], the SdH of $CeAgSb_2$ under pressure (1.2 kbar) for H//c showed a single SdH frequency of 25 T. As in Ref.14, we find the predominant frequency to be ~25T, as shown in FIG. 10a, and in high field (above 25 T), we observe another frequency (~300 T) below 2 K. We note the main oscillation frequency is significantly less than the lowest frequency (~40T) observed previously in the dHvA measurements.[2] We find that the amplitude of the 25 T oscillation is highly dependent on field orientation, where a tilt of only $7^0$ away from the *c*-axis causes a significant decrease in amplitude (FIG. 10b). FIG. 11 shows the effective cyclotron mass ($M_c$) for the 25 T orbit, which is ~ $3m_e$, extracted from the Lifshitz-Kosevich formula,[25] where we note that the effective mass, unlike the SdH amplitude, is not sensitive to the small change in angle. The cyclotron mass obtained in the previous dHvA measurement[2] is only $0.85m_e$ for the 41 T frequency. The Dingle temperature, which is a measure of scattering rate, obtained for field parallel to the c-axis and $7^0$ to the c-axis are 0.37 K and 0.66 K, respectively. This might imply that as the field is tilted away from the c-axis, the carrier scattering increases rapidly due to magnetic anisotropy, but as discussed below, other explanations are possible for this angular dependent attenuation of the SdH amplitude.

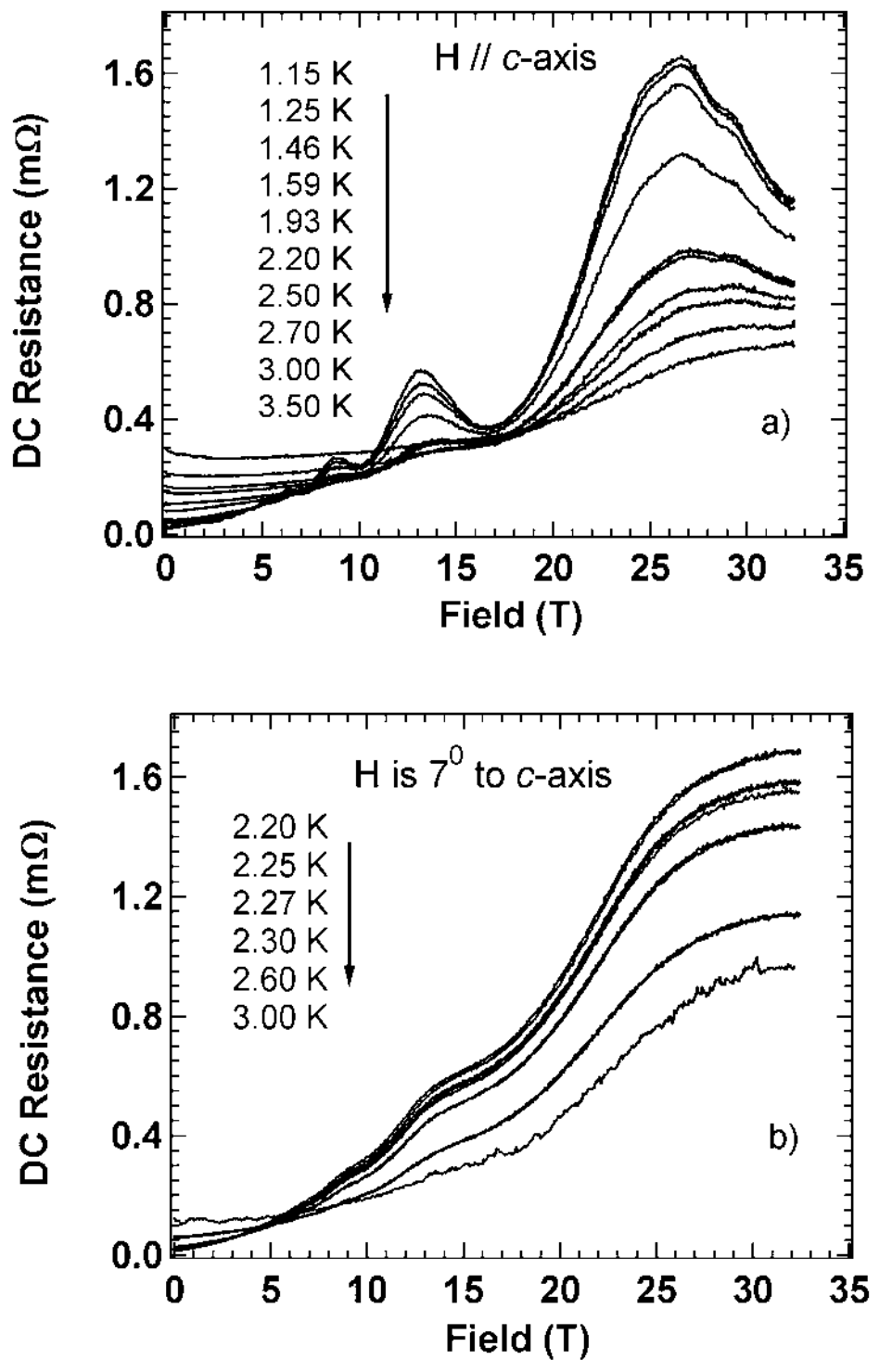

FIG. 10 Temperature dependence of Shubnikov-de Haas oscillations of $CeAgSb_2$ up to 32 T for two different angles.

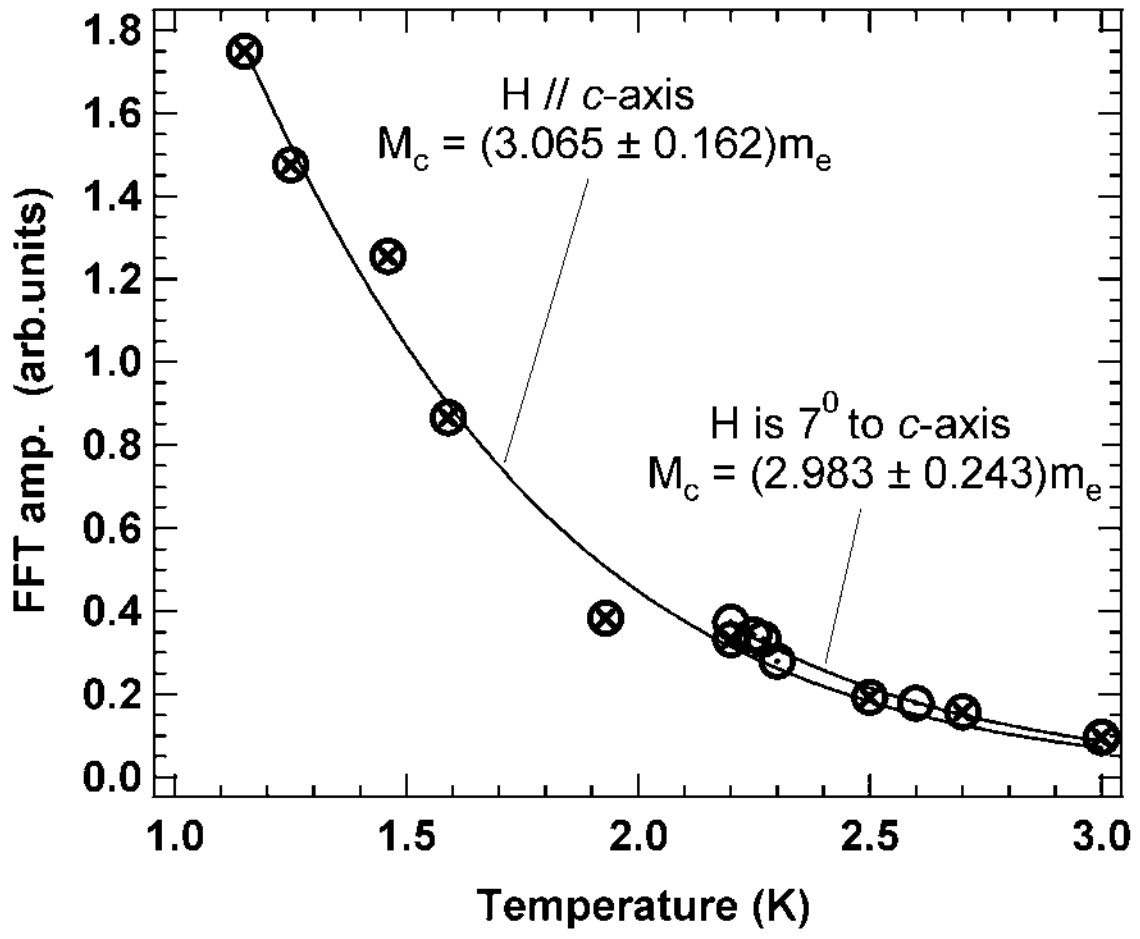

FIG. 11 FFT amplitude of the 25 T SdH oscillation in $CeAgSb_2$ vs. temperature for two different angles. The solid lines are fits to the Lifshitz-Kosevich effective mass expression.

The angular dependence of the SdH signal was studied more systematically at low temperature (~ 80 mK) to 18 T, as shown in FIG. 12a. We observe two different frequencies (~ 25 T and ~ 600 T) depending on the angle between the field and c-axis, as shown in FIG. 12b. The inset of FIG. 12b shows the amplitude of the 25T oscillation for different angles. The amplitude of the oscillation has a maximum around $95^0$, which is $5^0$ off from the *c*-axis, and very dramatically decreases away from $95^0$. The SdH oscillations below $80^0$ or above $115^0$ are very weak, and no SdH oscillations associated with the 25T frequency are observable outside this range.

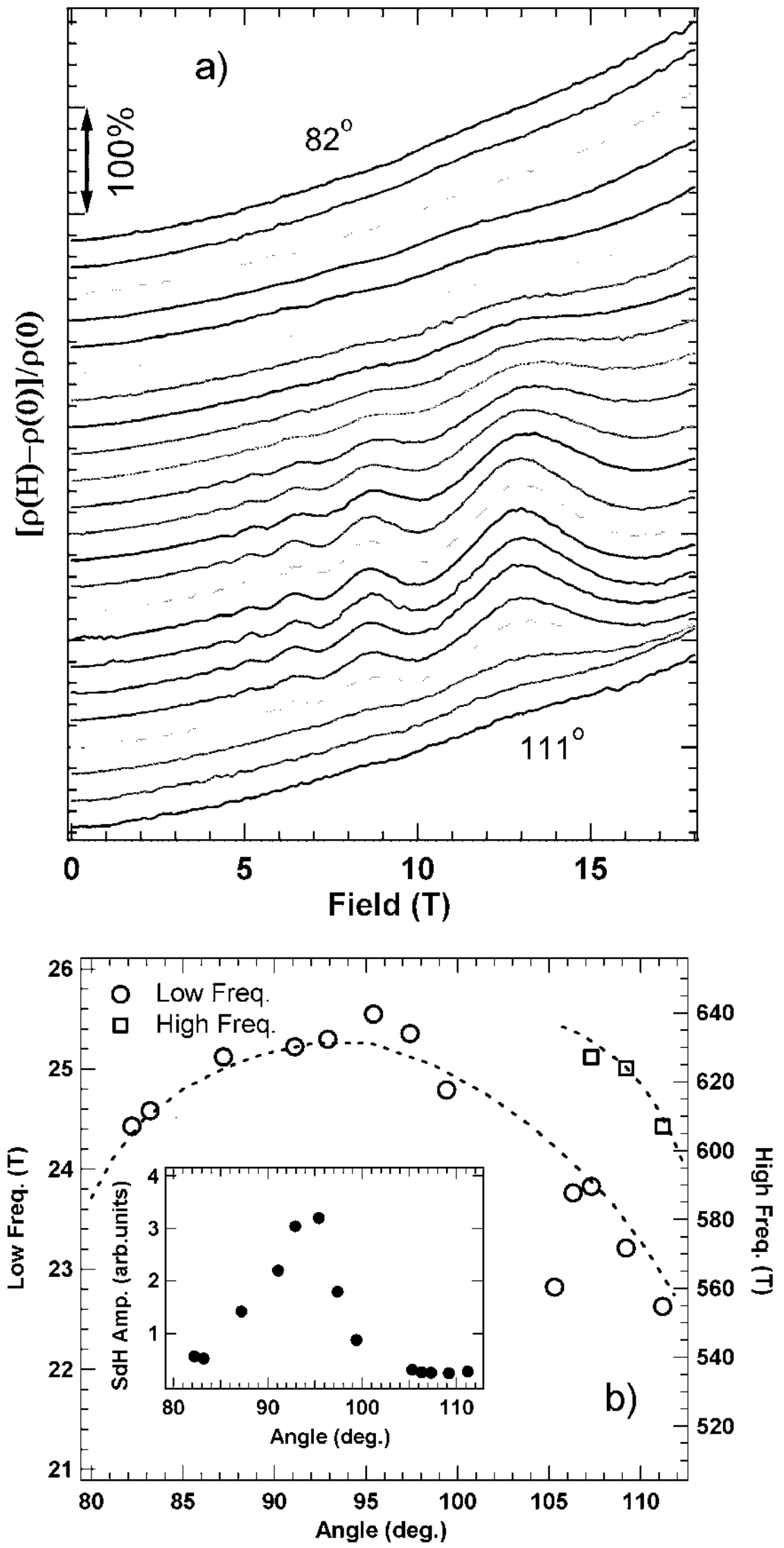

FIG. 12 a) The SdH oscillations of $CeAgSb_2$ for different angles at T~80 mK up to 18 T. b) SdH frequency vs. angle, dashed line is a guide to the eye. Inset: the amplitude of the oscillation vs. angle.

Both measurements indicate that the SdH amplitudes are very sensitive to magnetic field direction. The band calculation[2] suggests that the Fermi surface of $CeAgSb_2$ is similar to those of $LaAgSb_2$. The de Haas-van Alphen (dHvA) experiments indicate that the best model to use to explain the dHvA frequencies obtained in this compound is the model of band structure modified by the 4*f*-localized electron system.[2] This model is used to explain the topology of Fermi surface in Ce-compounds with magnetic order, such as $CeAl_2$ and $CeB_6$.[26,27] However, this can not describe all the frequencies observed in both SdH and dHvA measurements. More over, the decrease in the SdH frequency away from H//c is inconsistent with a simple two dimensional cylindrical Fermi surface directed along the c-axis. We believe the small frequency obtained in this experiment may come from a lens-shaped orbit due to overlapping Fermi surface sheets, perhaps at the zone boundary. In this case, especially if the main Fermi surface cylinders are corrugated, as the tilt angle increases, the cyclotron mass and frequency could be reduced,[28] as shown in FIG. 12. Moreover, the amplitude of oscillation might also be strongly affected by changes in the lens orbit topology as the angle moves away from $90^0$. The main Fermi surface, which has frequency of 11.2 kT, could not be obtained in this measurement. This might be due to the relatively large effective cyclotron mass, or to some other at present unknown fundamental difference between dHvA (a thermodynamic probe) and SdH (a transport probe) measurements in the present case.

**Conclusion**

The magnetic and electrical transport measurements we have performed on single crystalline $CeAgSb_2$ allow us to further advance a description of its magnetic ground state. To facilitate our description, we have derived and presented the theoretical expressions for electrical transport including magnon scattering for both the ferromagnetic and antiferromagnetic cases. At zero field the magnetic transition, based on this model, indicates that antiferromagnetic order appears below $T_O$~9.8 K. The magnetoresistance data, again compared with the magnon model, shows that at finite field antiferromagnetic order is present for in-plane field, and ferromagnetic order for field along the c-axis. One type of magnetic ground state that would lend itself to this anisotropy is a canted antiferromagnetic configuration in the basal plane. (A neutron scattering experiment on a single crystal would be very useful to test this possibility by determining in detail the spin alignments.) The molecular field parameter $\lambda$ extracted from CEF theory, which is related to the exchange interaction between the nearest neighbors, provides further evidence of the dependency of the magnetic ordering on field orientation. The magnetic H-T phase diagram we obtain from magnetotransport data is consistent with that found in previous magnetization and thermal expansion measurements[4]. A fit of the magnon model to the data yields a field dependent magnon energy gap, which is found to decrease with increasing field. Complementary to the field dependent magnon gap is the observation that the magnetoresistance changes sign at a temperature $T_m$ below $T_O$. $T_m$ is dependent on the magnetic field and its direction, and we present arguments to describe this effect.

Previous estimates of the Kondo temperature have varied widely. The Kondo temperature estimated from the specific heat coefficient is found to be 23 K, which is over a factor of two lower than those of obtained from preliminary inelastic neutron scattering and μSR measurements. However, a $T_K$ of 23 K is consistent with the resistivity measurement, which shows a coherence temperature of ~15 K, suggesting the Kondo temperature is of similar order. From the CEF analysis, we find $\Delta_1$ of 53K, which is similar to $T_K$ (60-80K) obtained from other measurements, and this may explain the conflicting estimates of $T_K$.

Remaining aspects of the $CeAgSb_2$ system that will require further investigation are the SdH results. We find that, in agreement with a previous preliminary study, that a small (25 T) orbit dominates the SdH signal, indicating significant differences between the SdH and previous dHvA results where no frequency below 40 T is observed. Moreover, unlike the dHvA, there seems to be a very strong attenuation of the SdH signal with field direction. Although SdH is sensitive to Stark-interference type orbits that have no thermodynamic weight, it is not clear why the 25 T is so evident, or if it does indeed arise from lens-like orbits from intersecting Fermi surface sections at the zone boundaries. It is also possible that the magnetic ground state may influence the electronic structure or the carrier mean free path in some unknown manner. Further high field SdH and dHvA comparative measurements are planned to explore these questions.

**Acknowledgements**

This research was sponsored by the National Nuclear Security Administration under the Stewardship Science Academic Alliances program through DOE Research Grant #DE-FG03-03NA00066. We would like to thank P. Schlottmann for fruitful discussions.